\documentclass[]{jfm}
\usepackage{amsmath,amssymb}
\usepackage{natbib}
\usepackage{labelfig,rotating}
\usepackage{epsf,graphicx}
\ifCUPmtlplainloaded \else
  \checkfont{eurm10}
  \iffontfound
    \IfFileExists{upmath.sty}
      {\typeout{^^JFound AMS Euler Roman fonts on the system,
                   using the 'upmath' package.^^J}%
       \usepackage{upmath}}
      {\typeout{^^JFound AMS Euler Roman fonts on the system, but you
                   dont seem to have the}%
       \typeout{'upmath' package installed. JFM.cls can take advantage
                 of these fonts,^^Jif you use 'upmath' package.^^J}%
      }
  \else
  \fi
\fi

\ifCUPmtlplainloaded \else
  \checkfont{msam10}
  \iffontfound
    \IfFileExists{amssymb.sty}
      {\typeout{^^JFound AMS Symbol fonts on the system, using the
                'amssymb' package.^^J}%
       \usepackage{amssymb}%

      }{}
  \fi
\fi
\ifCUPmtlplainloaded \else
  \IfFileExists{amsbsy.sty}
    {\typeout{^^JFound the 'amsbsy' package on the system, using it.^^J}
     \usepackage{amsbsy}}
    {\providecommand\boldsymbol[1]{\mbox{\boldmath $##1$}}}
\fi

\newcommand\Rey{\mbox{\textit{Re}}}  
\usepackage{color}

\title{Studying edge geometry in transiently turbulent shear flows}

\author[M. Chantry and T. M. Schneider]{Mathew Chantry$^1$ and Tobias M. Schneider$^2$}
\affiliation{$^1$School of Mathematics, University of Bristol, BS8 1TW, Bristol, UK\\[\affilskip] $^2$Max Planck Institute for Dynamics and Self-Organization, 37077 G{\"o}ttingen, Germany}

\begin{document}

\maketitle
\begin{abstract}
In linearly stable shear flows at moderate $\Rey$, turbulence spontaneously decays despite the existence of a codimension-one manifold, termed the edge, which separates decaying perturbations from those triggering turbulence. We statistically analyse the decay in plane Couette flow, quantify the breaking of self-sustaining feedback loops and demonstrate the existence of a whole continuum of possible decay paths. Drawing parallels with low-dimensional models and monitoring the location of the edge relative to decaying trajectories, we provide evidence that the edge of chaos does not separate state space globally. It is instead wrapped around the turbulence generating structures and not an independent dynamical structure but part of the chaotic saddle. Thereby, decaying trajectories need not cross the edge, but circumnavigate it while unwrapping from the turbulent saddle.  
\end{abstract}


\section{Introduction}
Turbulence in spatially confined linearly stable shear flows is observed to spontaneously decay for moderate Reynolds numbers. There has been extensive recent work studying this decay, specifically the statistical feature of turbulent lifetimes and how these vary with $\Rey$ \citep[see][]{faisst:life,hof:nature}. The result of this work suggests that in small domains there exists no critical Reynolds number for sustained turbulence, but that the mean lifetime increases with $\Rey$. In large domains such a critical Reynolds number appears to exist as expansion and splitting of localized turbulent patches becomes dominant over their decay \citep{avila2011}. We here focus not on the decay statistics but on the mechanism by which the turbulent dynamics spontaneously return to laminar flow and what the implications are for the global geometry of the system's state space.

The opposite process, the transition to turbulence triggered by finite amplitude bifurcations has as well been studied in detail. Initial perturbations which directly relaminarize are found to be separated from those that trigger turbulence by a codimension-one hypersurface embedded in the system's state space spanned by all instantaneous 3D velocity fields. In dynamical systems terms this hypersurface, now known as the \emph{edge of chaos} (or simply the {\emph{edge}}), generalises the basin boundary of the laminar fixed point for a situation where the turbulent state is not persistent. Using methods pioneered by \citet{itano} to track the dynamics within this edge, it was found to be -- at least locally -- formed by the stable manifold of a relative local attractor termed edge state. These edge states are invariant solution of the Navier-Stokes equations with a single unstable direction so that they attract dynamics within the edge. Tracking the dynamics within the hyper-surface between laminar and turbulent flow allowed discovery of several previously unknown invariant solutions originally in small domains \citep[see][]{skufca:edge,kerswell2007recurrence,schneider:turb,duguet2008,viswanath2008,kim2008characterizing} and in larger domains leading to spatially localized solutions \citep[see][]{schneider2010,avila2013}.
\begin{figure}
\begin{center}
\includegraphics[scale=0.3]{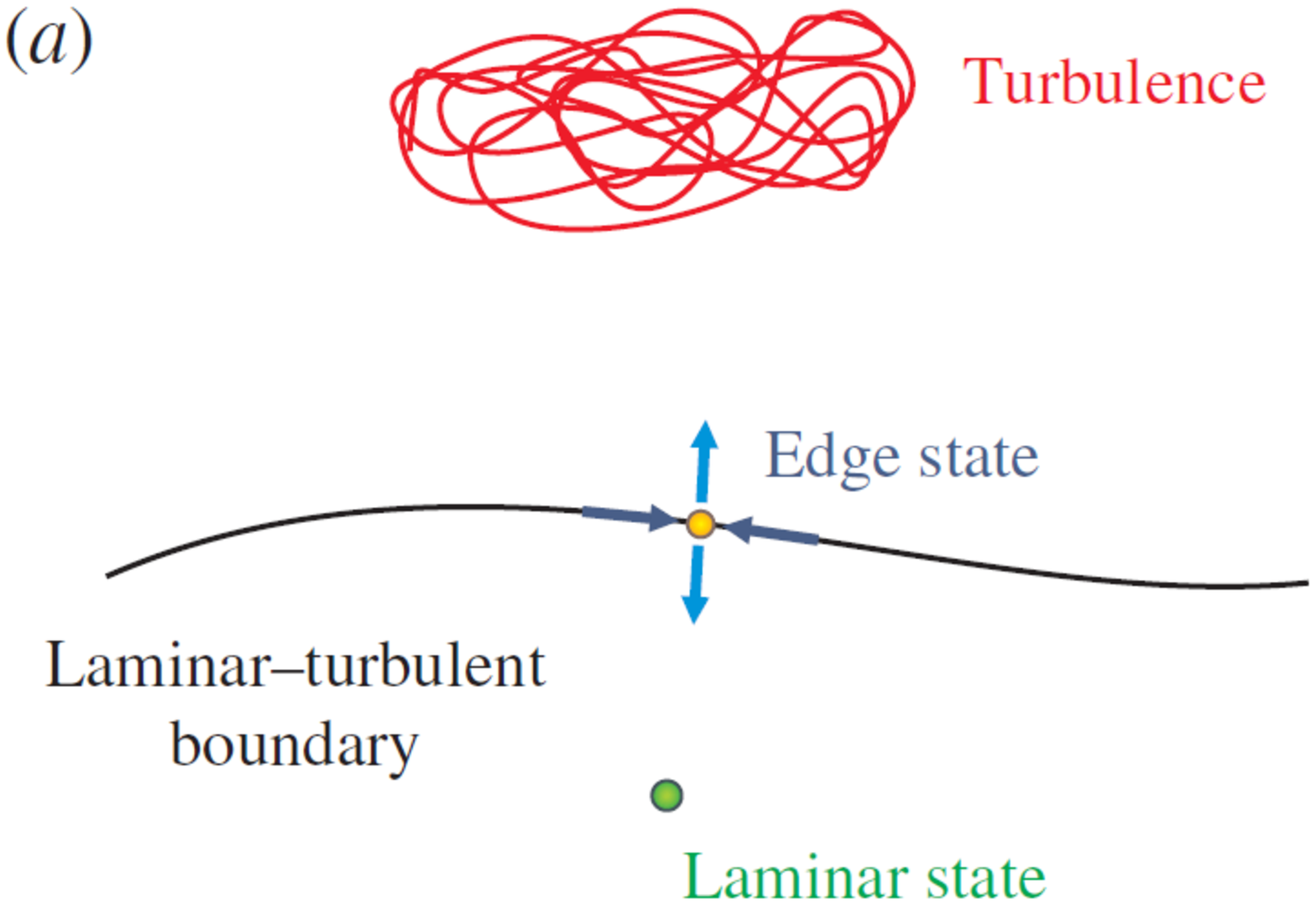}
\includegraphics[scale=0.3]{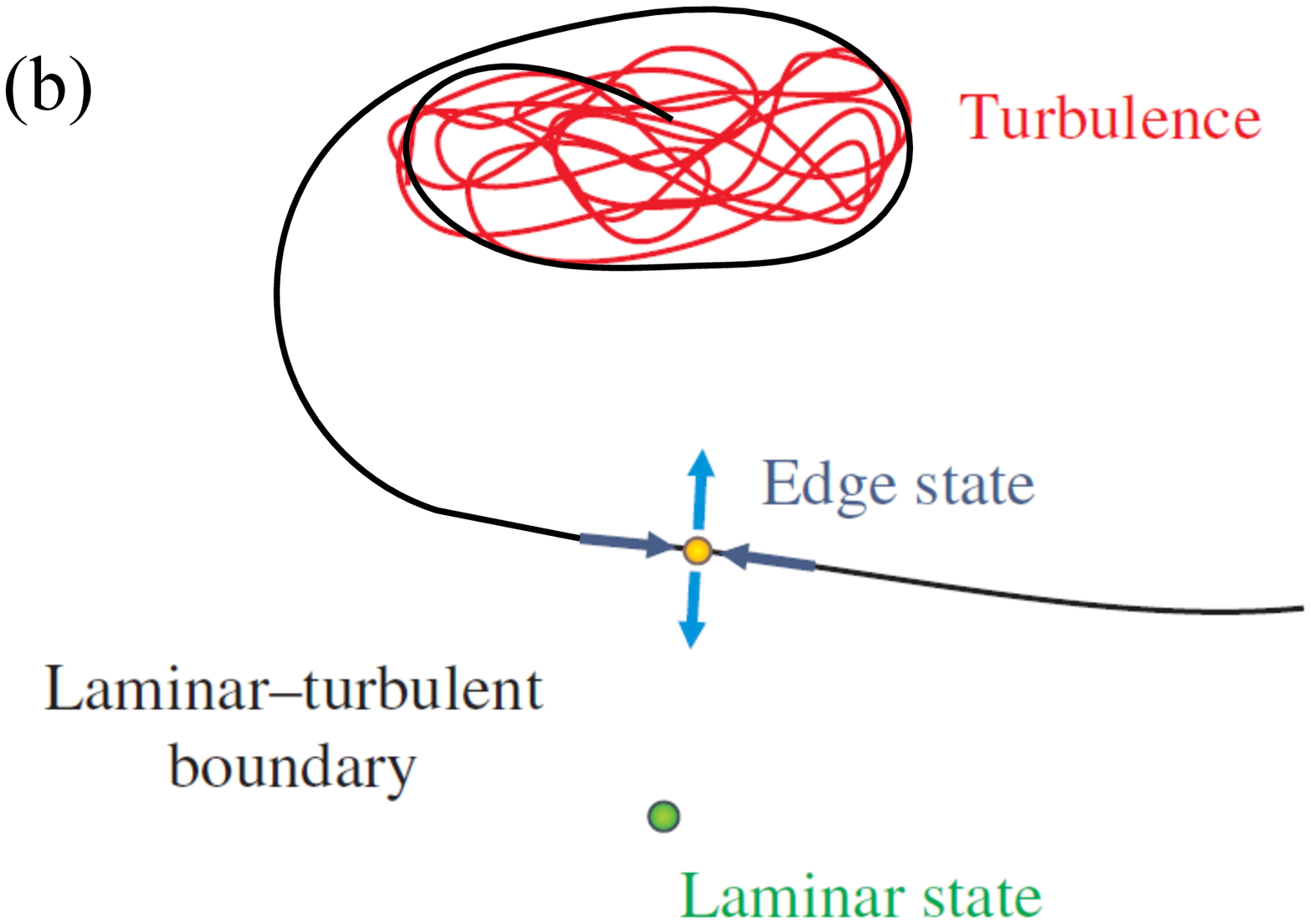}
\caption{Two models for the state space of transient turbulence with laminar flow (green) and the turbulent saddle (red). Between these lies the edge state (yellow) and its stable manifold the edge of chaos (black), a codimension-one hypersurface. (a) The classical view, with this particular cartoon from \citet{khapko13}, where state space is divided by the manifold extending to infinity leaving no clear route for decaying turbulence. (b) An adaptation inspired by \citet{lebovitz12} where the edge is not dynamically separate from the turbulent saddle but wrapped up into it allowing an unwinding route to the laminar state.
}
\label{fig:khapko}
\end{center}
\end{figure}

The common view of the systems state space as depicted in a recent illustration by \citet{khapko13} is that of a chaotic saddle supporting turbulence in one region, the attracting laminar fixed point in another domain and the edge separating those two domains. Thus, there are three distinct dynamical objects governing state space: the turbulent saddle, the laminar fixed point and the edge with the embedded edge state in between. The depicted global geometry, though compatible with observations at the transition, is challenged by the observed spontaneous decay of turbulent fluctuations. A decaying trajectory dynamically connects the turbulent saddle with the laminar fixed point but the edge is formed by a codimension-one manifold which can generally not be crossed. Moreover, being a manifold, the edge in general does not have a boundary at which it terminates and is therefore assumed to extend to infinity in figure \ref{fig:khapko}(a). Consequently within the common picture a decaying trajectory cannot circumnavigate the separating edge. Thus, there remain questions about the global topology of the edge, how it allows for a decaying trajectory to ``migrate'' from the turbulent to the laminar side and along which paths such a decay occurs. It has been speculated that there are specific isolated points associated with embedded invariant solutions that facilitate the piercing of the edge. In a smooth manifold, those points generally do not exists and can thus only be rare. Consequently, only a unique (or a small number of discrete) decay paths remains possible. Results in pipe flow were interpreted as support that the edge state itself is one of these portals and thus not only governs the transition to but also the decay from turbulence \citep[e.g.][]{delozar12}.

In this paper we aim to elucidate the somewhat murky picture through a detailed study of decaying trajectories in a plane Couette cell, where there is a well studied codimension-one edge yet turbulence spontaneously decays. We describe the evolution of the decaying flow fields and demonstrate that there are not a small number of discrete decay channels but a whole continuum of possible paths, which is incompatible with the state space structure discussed above. Instead all data is compatible with an alternative geometry which was suggested by an extremely simplified 2D ODE model inspired by shear flows. In this model, studied by \citet{lebovitz12}, the edge does not extend to infinity in all directions but is  ``wrapped around'' the non-laminar state. The emerging picture found is the one depicted in figure \ref{fig:khapko}(b) where the edge is actually not separated from but part of the turbulent saddle. Thus, it no longer bounds the laminar state away from turbulence and decaying trajectories need not cross the edge. 

\section{Statistics of decay}

In the following we quantitatively describe the decay from turbulence and demonstrate that the statistics are compatible with the Lebovitz picture for a high-dimensional flow system. While for fully resolved flow simulations, with at least $10^5$ degrees of freedom, studying the global geometry of the edge manifold in all dimensions remains impossible despite modern computational tools, we analyse the dynamics of initial conditions along sections of state space. The variation observed along those sections shows precisely the signature expected from the low-dimensional models thus further supporting the conclusion, that the correct model for moderate Reynolds number shear flows is not an infinitely extended edge but an edge dynamically connected to the turbulent saddle.

We study plane Couette flow in a periodically continued domain of downstream length $L_x = 4 \pi$ and spanwise extent of $L_z=2 \pi$, where as usual lengths are measured in terms of half the plate separation $h$. Decay in small domains has been extensively studied and clearly shows exponential lifetime distributions
\citep{schneider:decay}. For Reynolds number $\Rey=380$, where  $\Rey:=U h / \nu$, with $U$ is the wall speed, $h$ is half the wall separation and $\nu$ the kinematic viscosity, the edge is formed by the stable manifold of a single fixed point \citep{schneider:lam}, a version of the historically first invariant exact solution identified in plane Couette flow by \citet{nagata90}. Thus, we can statistically study decaying trajectories in the presence of codimension-one edge manifold and investigate to which extent discrete decay paths are observed.

We aim at quantifying the physical processes initiating turbulence decay as well as their statistical variation. To this end, an ensemble of decaying trajectories is constructed numerically. Technically, we use Channelflow, a spectral software package \citep[for details see][]{gibson08,channelweb}. We run our simulations at a spatial resolution of $64 \times 36 \times 64$ modes corresponding to over $10^5$ degrees of freedom. Two hundred transients starting from different initial conditions are followed until decay for $\Rey=380$, where characteristic lifetimes are on the order of $1000$ time units. Due to the characteristic randomness of the decay, comparing different decays and studying the processes initiating the decay requires measurement of time relative to when the flow has decayed. Thus, we align all trajectories by remapping time according to   $t^\star := t_{lam} - t$ , where the relaminarization  time  $t_{lam}$ is the smallest $t$ such that $\| \mathbf{u}\left(t\right) \| < \epsilon = 0.005$ with $\mathbf{u}$ the deviation from laminar flow. The criterion for detecting decay is chosen such that the subsequent evolution is guaranteed to follow the exponential approach governed by the Navier-Stokes equations linearised around the laminar state. Neither the parameters nor the specific form of the chosen criterion affect our results.

The decay of turbulent motion is associated with the breaking of the feedback mechanisms that otherwise sustain the dynamics. This feedback loop analysed in detail by \citet{waleffe97} involves predominantly downstream oriented vortices (``rolls'') which generate growing streaks. These streaks undergo a 3D instability which via nonlinear interaction feeds back onto the rolls. To quantify these interactions we thus monitor the $L^2$ norms not only  of the full perturbation field, $\mathbf{u}$, but also the downstream velocity component, $u$, indicating the strength of streaks, the downstream vorticity $\omega_x$, capturing downstream vortices and of the three-dimensional velocity component, $\mathbf{u'} =\mathbf{u} - \langle\mathbf{u}\rangle_x$.

\begin{figure}
\begin{center}
\SetLabels
(-0.05*0.85) {\small (a)} \\
(-0.03*0.55){\small $\|u\|$} \\ 
(0.52*0.03){\small $t^*$} \\ 
\endSetLabels 
\strut\AffixLabels{\includegraphics[width=60mm]{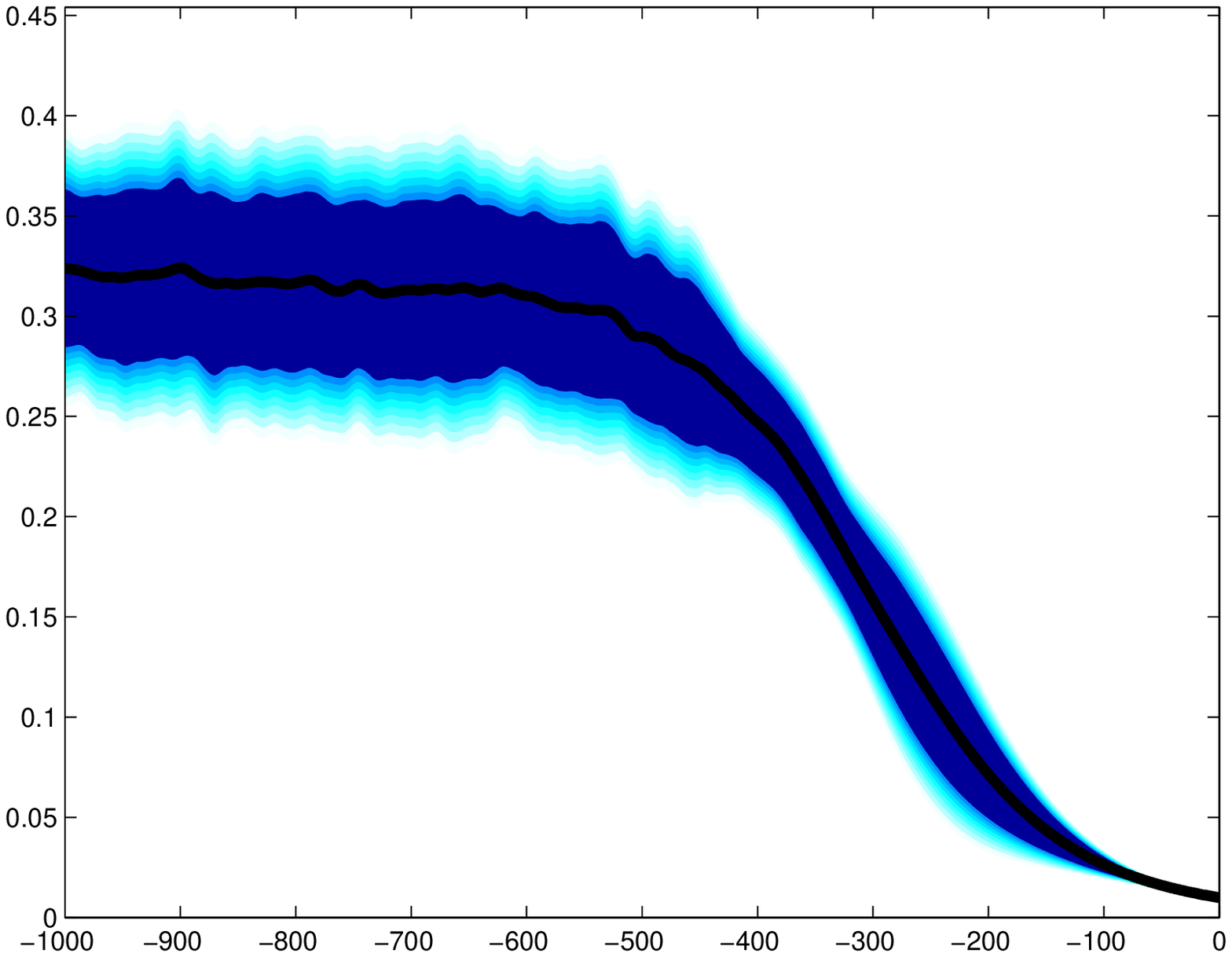}}
\hspace{5mm}\SetLabels
(-0.05*0.85) {\small (b)} \\
(-0.03*0.55){\small $\|\boldsymbol\omega_x\|$} \\ 
(0.52*0.03){\small $t^*$} \\ 
\endSetLabels 
\leavevmode
\strut\AffixLabels{\includegraphics[width=60mm]{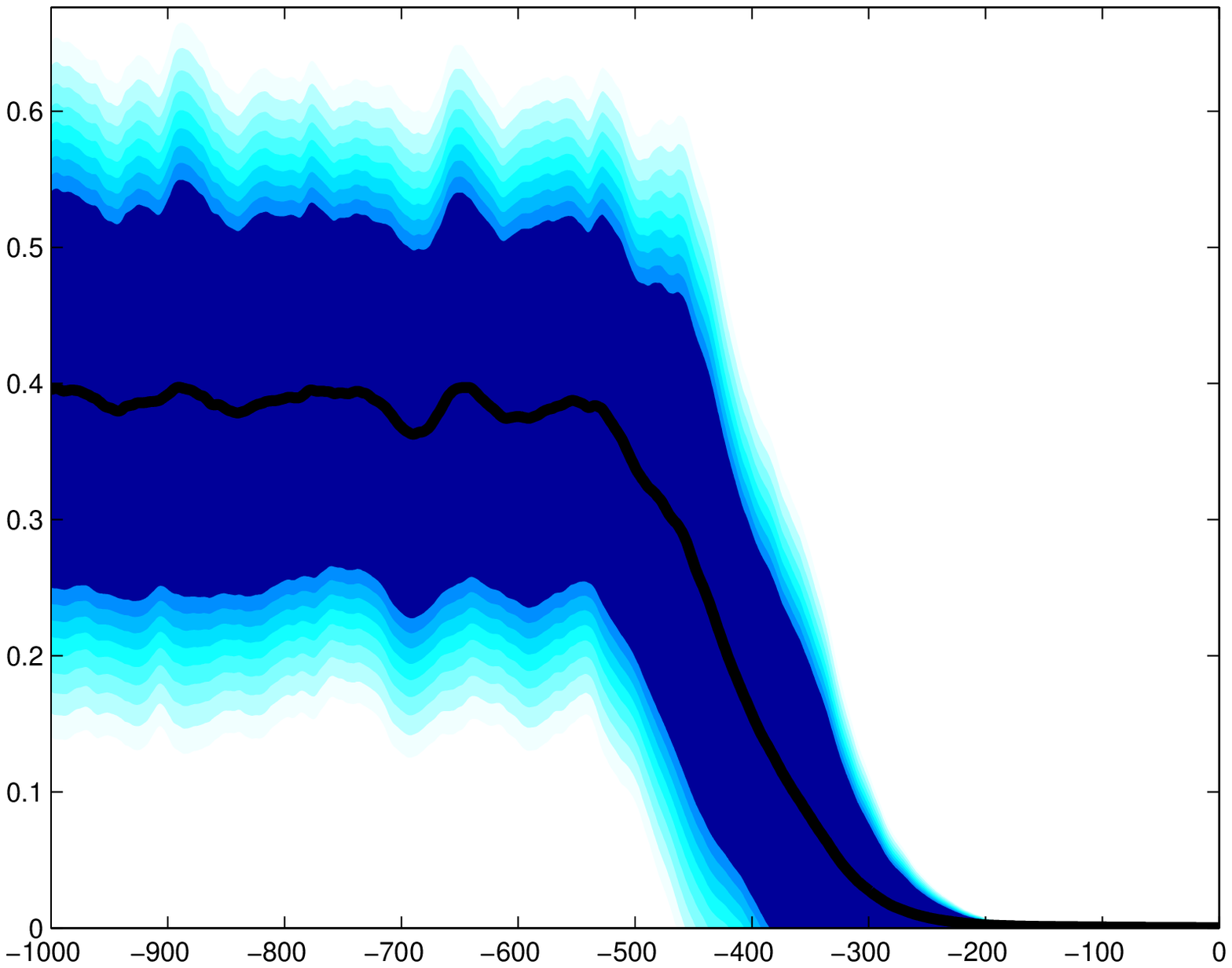}}
\caption{Mean decay from turbulence with trajectories aligned by decay time, the first standard deviation is plotted in dark blue varying to light blue over the second. We plot the $L^2$ norms of (a) the downstream velocity, $u$, representing the streaks, (b) the downstream vorticity, $\omega_x$, representing the rolls. There are two phases of decay, the first where rolls (and instabilities which are not plotted) rapidly decay and then the viscous exponential decay of the streaks.
}
\label{fig:380}
\end{center}
\end{figure}

Figure \ref{fig:380} shows two of the ensemble averaged measures discussed above as a function of $t^\star$ plotted with two standard deviations about the mean. The decay starts at $t^\star \approx -500$, that is 500 time units before the close neighbourhood of the laminar state is reached. While the evolution indicates both measures start decaying simultaneously, the quickest decay within about $250$ units is observed for the downstream vorticity. The three-dimensional component of the velocity decays quickly together with the downstream vorticity and the full perturbation field follows the slow streak dynamics. The behaviour of the three-dimensional velocity mirrors that of the downstream vorticity and so is not plotted. Consequently, during a first step of the decay, downstream vortices disappear and the field loses its downstream dependence. By $t^\star\approx -250$, the energy contained within the rolls and instability parts of the flow has decreased to a negligible amount. At this point the decay enters the final regime where the streaks are no longer forced but viscously damped away. Consequently the energy decreases exponentially with a decay rate set by the spatial structure of the streaks remaining once the flow has lost its vortices.

\begin{figure}
\begin{center}
\SetLabels 
(-0.03*0.55){\normalsize $\|\mathbf{u}\|$} \\
(0.55*-0.03){\normalsize $t^\star$} \\
\endSetLabels 
\leavevmode
\strut\AffixLabels{\includegraphics[scale=0.95]{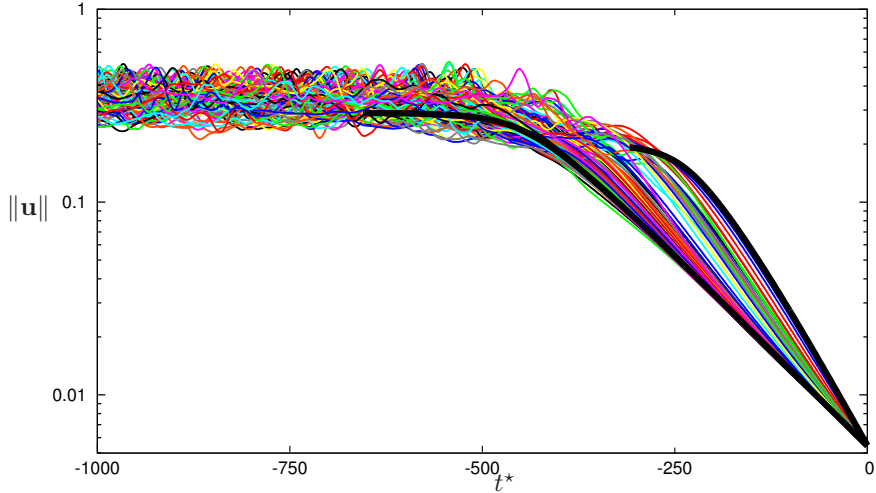}}
\caption{Decay of individual trajectories (coloured lines) when aligned by decay time. Note the continuum of exponential decay rates incompatible with specific crossing points in the edge. Decay is bracketed by decay from two solutions from the ``Nagata'' family embedded within the edge, where the faster decaying solution is the edge state.}
\label{fig:decay}
\end{center}
\end{figure}

Besides the two steps of the decay, the other striking feature of the data in figure \ref{fig:380} is the large statistical variation between different decaying trajectories. As a result the standard deviation remains large for almost the entirety of the decay. This suggests the absence of a preferred decay path. If there were distinct structures at which the trajectory traverses the edge the further evolution towards laminar flow would follow a unique route. Consequently the statistical variation observed in turbulence should collapse when the decay commences. Our data clearly does not support this scenario. Instead we observe a whole continuum of possible decay rates during the second exponential phase of the decay (see figure \ref{fig:decay}). 

The observed exponential decay of the predominantly $x$-independent streak patterns is controlled by the viscous term of the Navier-Stokes equations and thereby directly related to the length-scales of the pattern. The rate depends predominantly upon the relative amplitude  of the first and second Fourier modes in the spanwise direction $z$. In the downstream flow these represent one and two pairs of the fast-slow streaks, generated by the now-decayed roll structure. This point is illustrated by the perturbed dynamics of two fixed point solutions from the ``Nagata'' family embedded in the edge. The solutions have small downstream variation and serve as examples of different streak structures. One is the edge state which has wavelength in $z$ of $\pi$ and therefore two pairs of downstream streaks and a second solution with a wavelength of $2\pi$ in $z$ but the same wavelength in $x$ and consequently one pair of streaks. Perturbing the fixed points in the direction of laminar flow (to the ``laminar side'' of the edge) provides decaying trajectories (black lines in figure \ref{fig:decay}) to compare with the ensemble of decaying trajectories. Decay from the edge solution forms a upper bound for decay rate of trajectories, while decay from the second fixed point solution is close to a lower bound for the decay rate. Thus, all trajectories exit the turbulent saddle with at least some energy in the mode corresponding to the largest wavelength the domain supports and which is twice the wavelength of the edge state. No significantly faster decay than that of the edge state is observed, suggesting that streak patterns with predominantly higher number of streaks are subdominant. This figure demonstrates a smooth range of decay rates reiterating the lack of evidence for a small number of decay paths. There is clearly no preferred decay rate and there is specifically no evidence for preferred close passage of the edge state, with very few trajectories demonstrating similar decay rates. This observation is robust to changing the metric used for temporal alignment of the decay. 

In summary, the observed  continuum of decay paths is not compatible with a small number of discrete portals that allow traversal of the edge of chaos and spontaneous decay can thus not be reconciled with an edge manifold separating state space. All data however remains compatible with the alternative scenario of an edge wrapped around the turbulent saddle. To further explore to what extent the alternative scenario applies to the high-dimensional shear flow state space, we compare observations with those in the ODE model by \citet{lebovitz12} in which the scenario takes the simplest form and due to being only two dimensional allows for a complete graphical representation of state space. 

\section{Model comparisons}

\begin{figure}
\begin{center}
\includegraphics[scale=0.55]{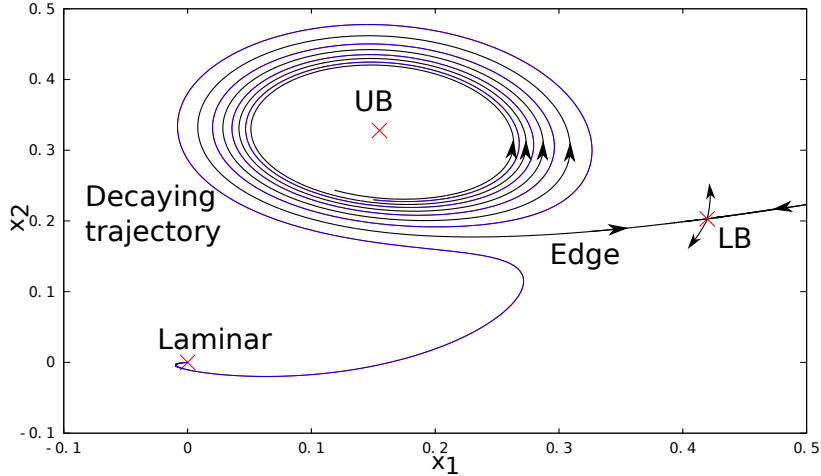}
\caption{State space for the 2D model from \citet{lebovitz12} for parameter $R>2.5$. The laminar state and the two other steady states are denoted by red crosses. The unstable UB solution is the surrogate for turbulence,
LB is the edge state and its stable manifold is the edge. The edge spirals infinitely many times around the UB forming part of its unstable manifold and providing a route back to the laminar state. For visualization this inner portion of the edge is not plotted. A typical ``decaying'' trajectory is plotted in blue.}
\label{fig:model}
\end{center}
\end{figure}

The two-dimensional model introduced by \citet{lebovitz12}, the complete dynamics of which are illustrated in figure \ref{fig:model}, is the simplest of a hierarchy of models of increasing dimensionality \citep[][]{lebovitz09,lebovitz13} which share features of the edge geometry where the edge is wrapped up and around the non-laminar dynamics. The model has a stable ``laminar'' fixed point solution plus two other unstable solutions labelled LB (lower branch) and UB (upper branch). In the model, the UB fixed point represents ``turbulence'', while LB is the edge state. The stable manifold of LB forms the edge (black curve) and wraps infinitely many times around UB and is part of its unstable manifold, but for visualization purposes this connection is not shown. Thus, the edge does not extend to infinity but dynamically connects to the surrogate for turbulence. Still, the outermost ring of the edge manifold displays the defining edge properties with conditions on one side returning to the laminar state, while those on the other explore non-trivial dynamics before returning to the laminar state. This geometry introduces a range of routes for trajectories to return to the laminar state from the UB state. Considering the example of a decaying trajectory in figure \ref{fig:model} (blue curve) we see consistency with our statistical observations from the full system: trajectories do not exit the spiral at a specific point, they do not need to closely approach the edge state, the final decay rate close to the laminar fixed point varies continuously and depends upon the specific route taken.

\begin{figure}
\begin{center}
\SetLabels 
(0.03*0.9){\normalsize (a)} \\
\endSetLabels 
\leavevmode
\strut\AffixLabels{\includegraphics[width=0.7\textwidth]{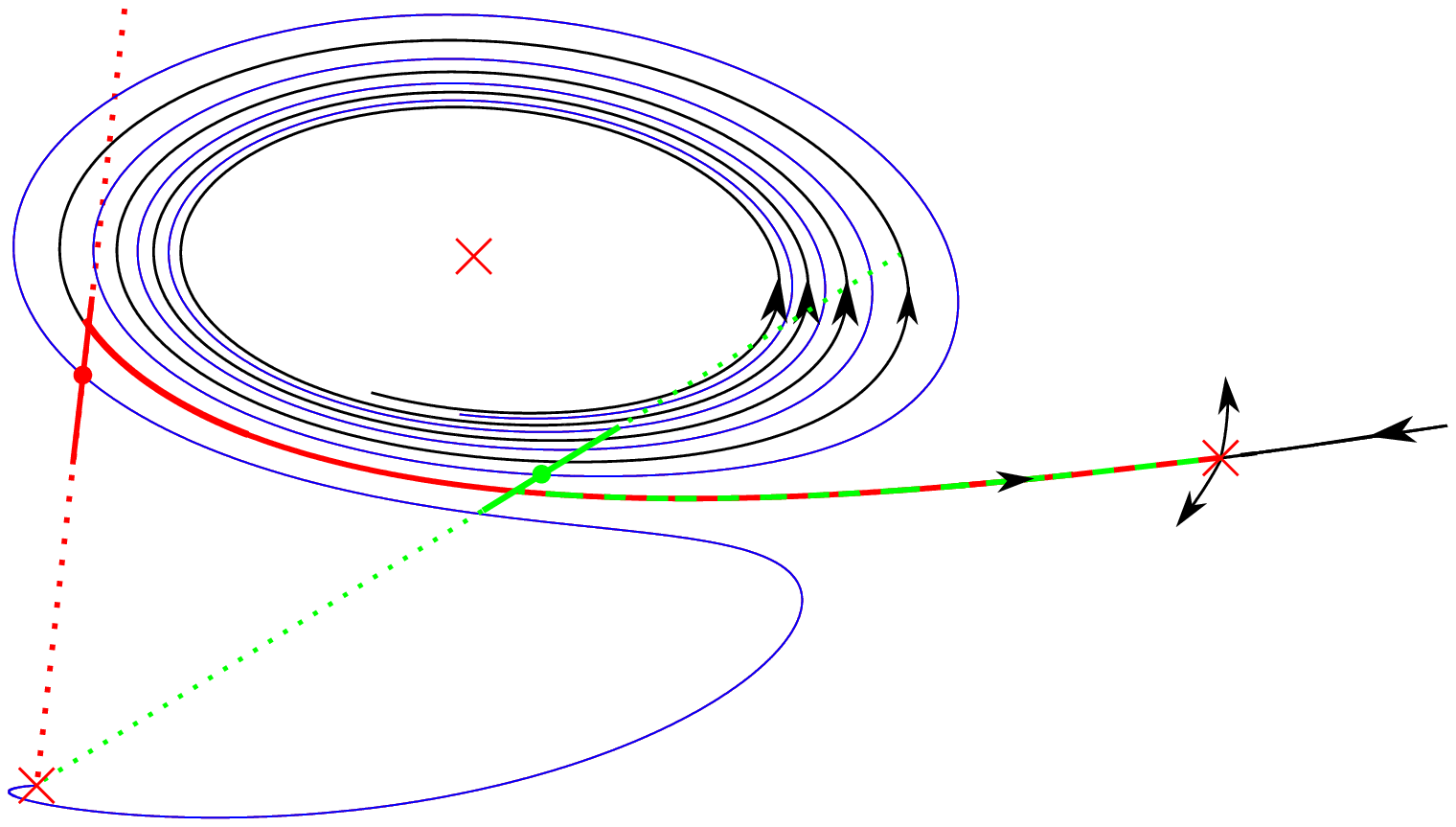}}
\vspace{3mm}

\SetLabels 
(0.52*1.0){\large Model} \\
(-0.01*0.9){\normalsize (b)} \\
(-0.00*0.40){\normalsize \rotatebox{90}{Lifetime} } \\
(0.52*-0.02){\normalsize $\lambda$} \\
\endSetLabels 
\leavevmode
\strut\AffixLabels{\includegraphics[width=0.49\textwidth]{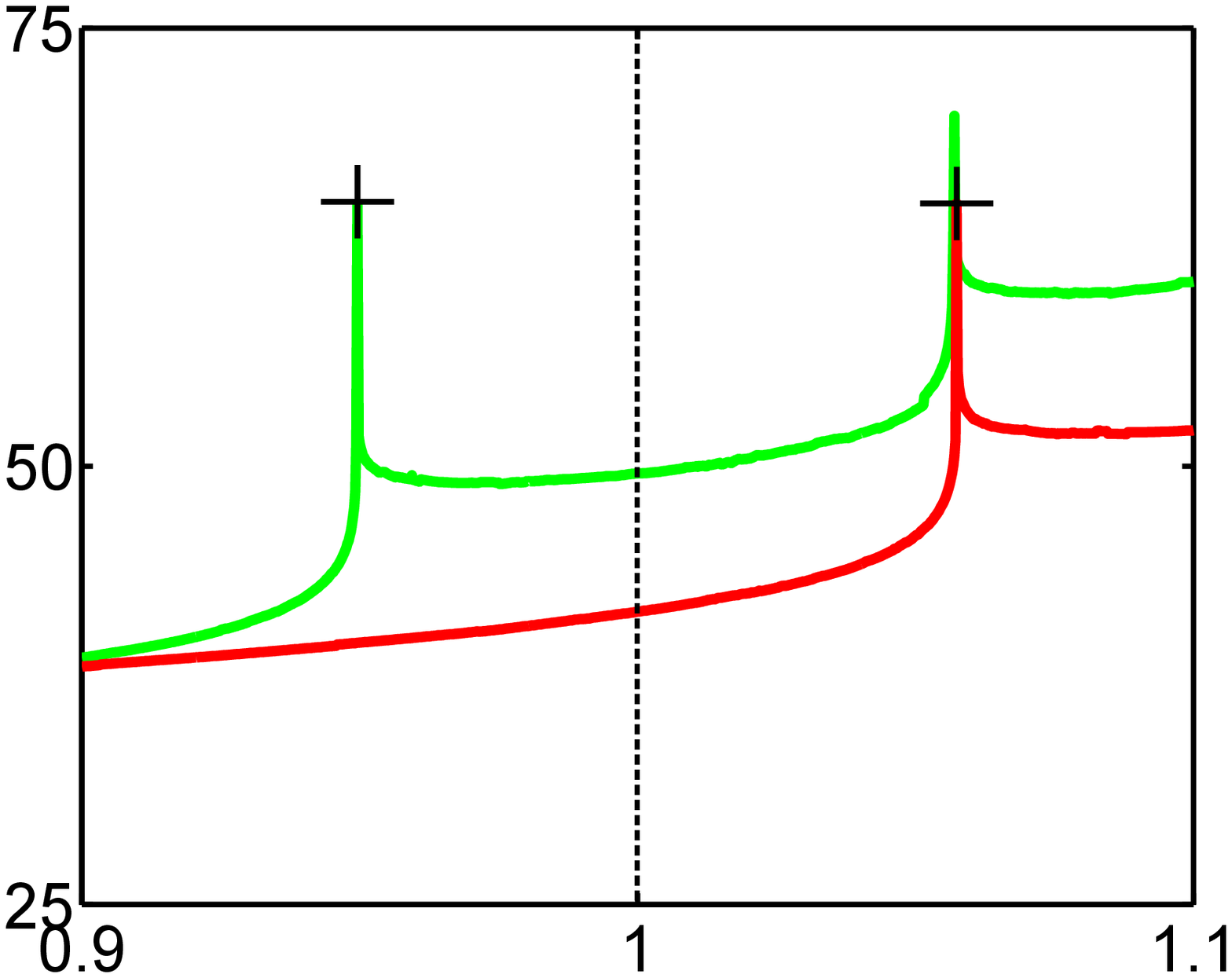}}
\SetLabels 
(0.52*1.0){\large pCf} \\
(-0.01*0.9){\normalsize (c)} \\
(-0.03*0.40){\normalsize \rotatebox{90}{Lifetime} } \\
(0.52*-0.02){\normalsize $\lambda$} \\
\endSetLabels 
\leavevmode
\strut\AffixLabels{\includegraphics[width=0.49\textwidth]{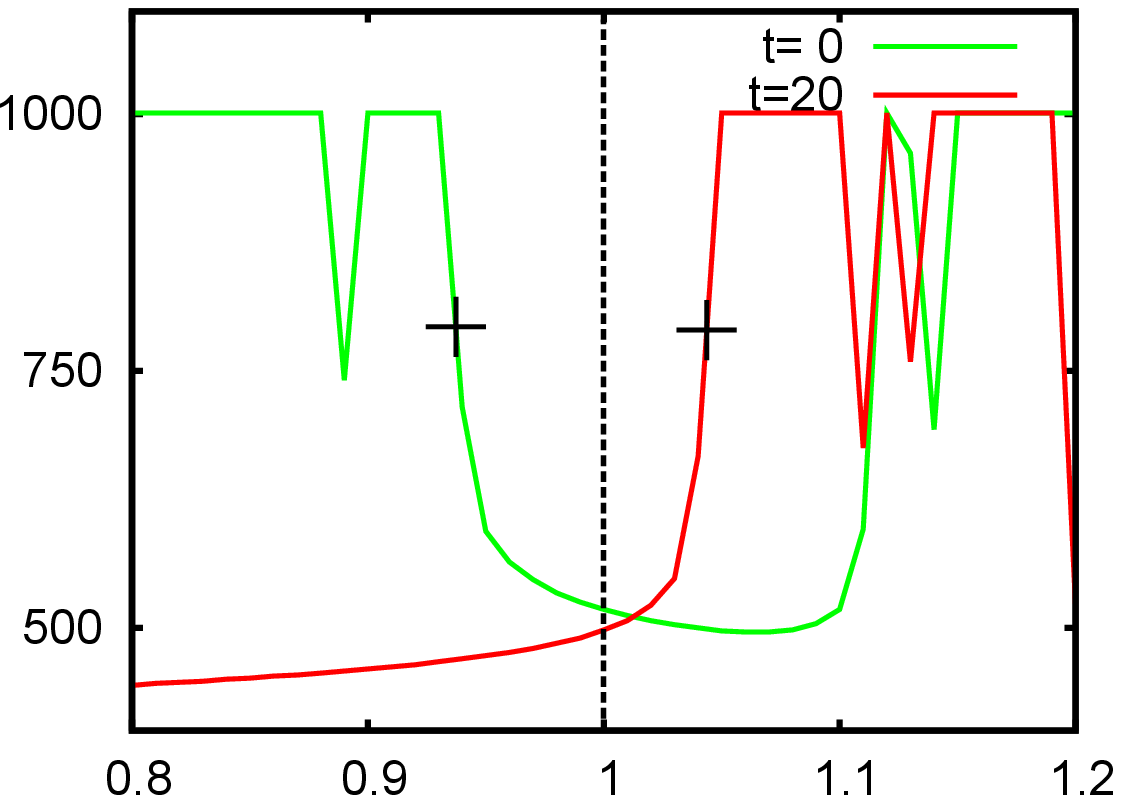}}

\SetLabels 
(0.0*0.9){\normalsize (d)} \\
(-0.03*0.55){\normalsize $\|\mathbf{x}\|$} \\
(0.52*-0.02){\normalsize $t$} \\
\endSetLabels 
\leavevmode
\strut\AffixLabels{\includegraphics[width=0.49\textwidth]{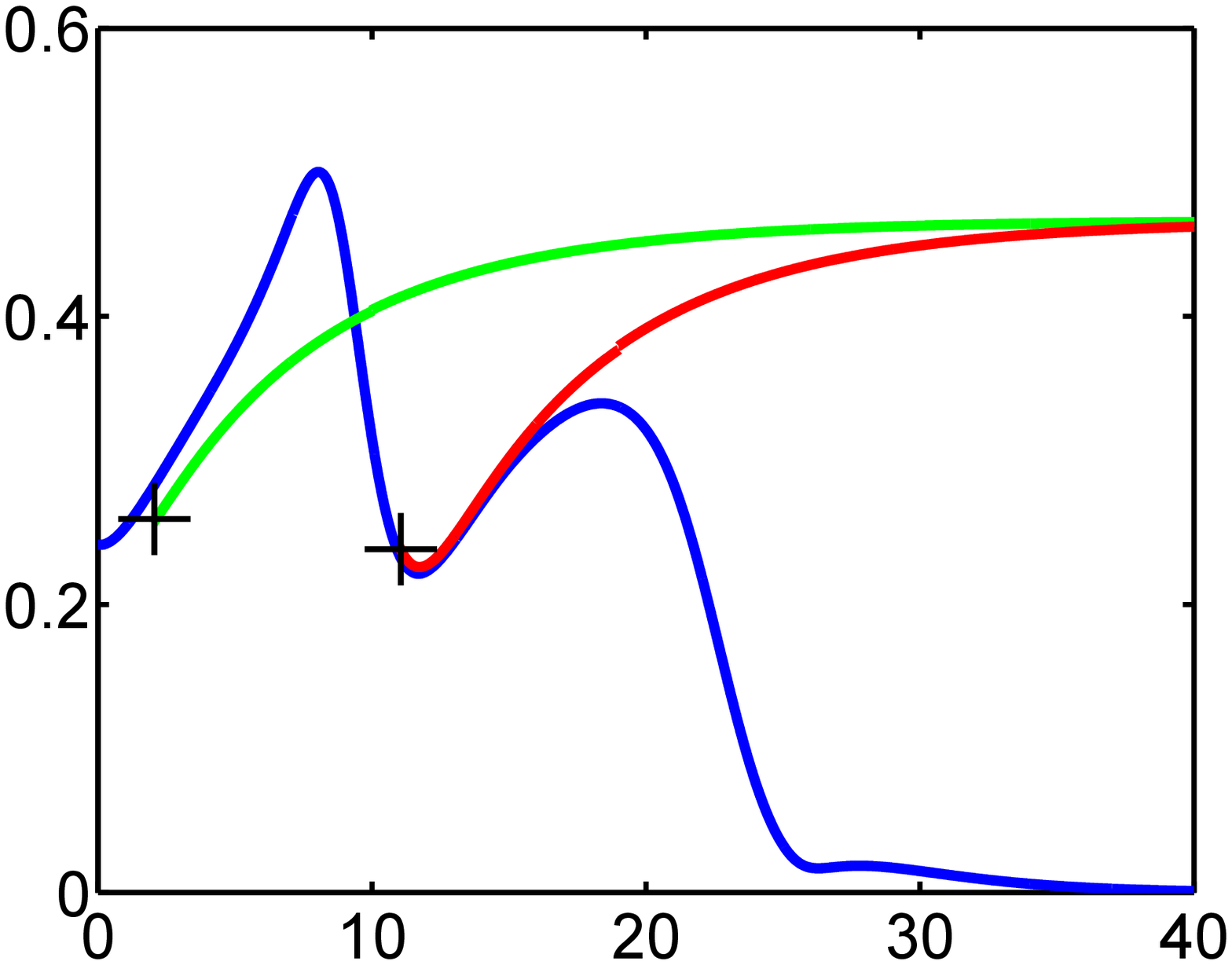}}
\SetLabels 
(0.0*0.9){\normalsize (e)} \\
(-0.03*0.55){\normalsize $\|\mathbf{u}\|$} \\
(0.52*-0.02){\normalsize $t$} \\
\endSetLabels 
\leavevmode
\strut\AffixLabels{\includegraphics[width=0.49\textwidth]{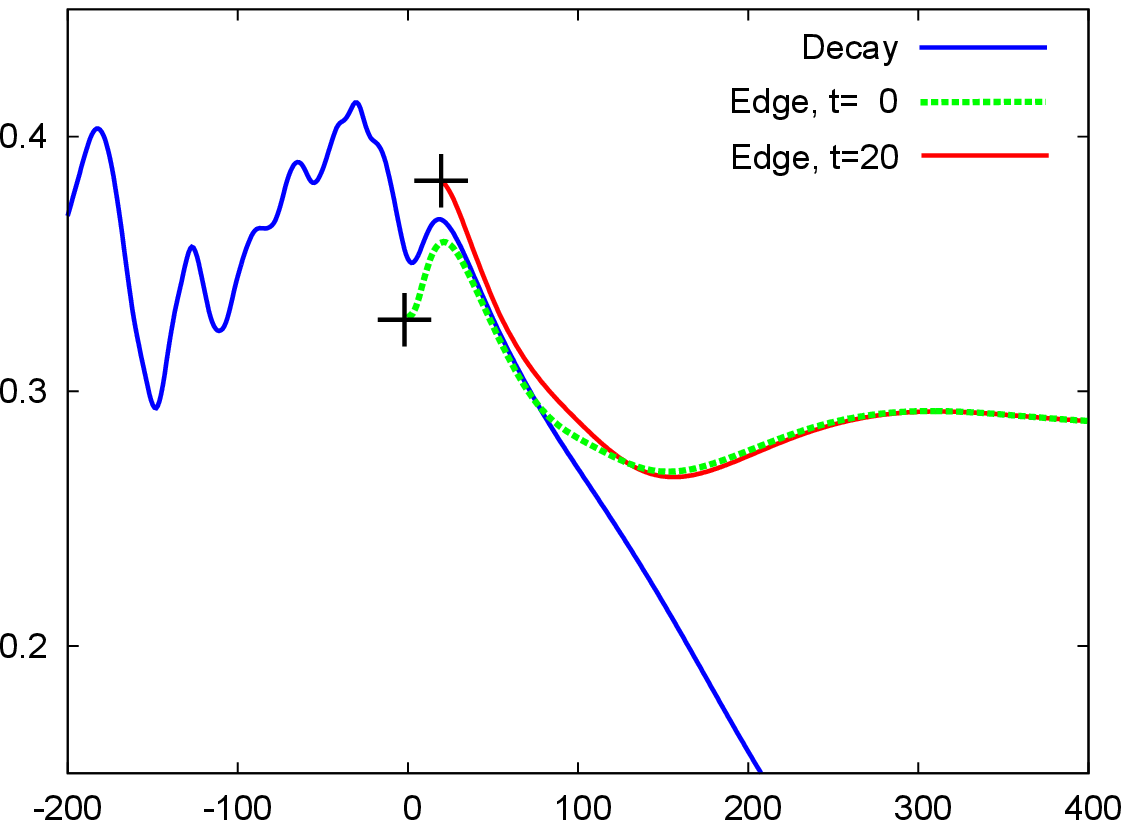}}
\caption{Evidence for shared geometry features between the model and pCf with frame by frame comparison between the two.
Frame (a): State space for the reduced model illustrating the techniques used to locate the edge (black line) location relative to the decaying trajectory (blue line). At two points along the trajectory (indicated by dots) the condition is rescaled along the dotted lines connecting the point to the laminar state. Early in decay the edge pieces bracket decay (green point) whereas later the edge lies only on one side of the trajectory (red point). 
Frames (b,c): Lifetime against rescaling parameter $\lambda$ for two points along a decaying trajectory in the model (frame (b)) and pCf (frame (c)). In the model peaks in lifetime denote the edge. In pCf transitions between long (turbulent) and short (relaminarizing) lifetimes indicate the edge. 
Frames (d,e): Edge pieces in both systems are tracked leading to shared dynamics. Edges indicated by crosses in frames (b,c) are tracked in green (early in decay) and red (later in decay). In both systems the edge pieces both early and later in the decay lead to shared dynamics. 
}
\label{fig:edgeround}
\end{center}
\end{figure}

Thus, all statistical data supports the interpretation that this low-dimensional model exemplifies the global state space structure of moderate $\Rey$ shear flows. To further support this interpretation we consider the topology of the edge, namely the dynamical connection to the turbulent state which allows decay around the edge. Despite the low-dimensionality we speculate that the geometry, with the characteristic of being wrapped around the turbulence, is a defining feature and search for evidence of a similar structure in plane Couette flow. To this end we follow a decaying trajectory in pCf and locate the edge (black in figure \ref{fig:edgeround}(a)) relative to the trajectory. 

In the model state space, which here serves as a suggested schematic of pCf, the edge can be completely visualized. To compare the model and pCf we carry out equivalent processes which are plotted in \ref{fig:edgeround}(b,d) and \ref{fig:edgeround}(c,e), respecitively. In frame (b) we plot the lifetime of initial conditions generated by rescaling points along the decaying trajectory  as $\mathbf{x}_{\lambda} \left(t_0\right) = \lambda \mathbf{x}\left(t_0\right)$. Early in the decay (green), the edge lies on both sides of the decaying trajectory. The lifetimes diverge for $\lambda$ values both larger and smaller than 1 when the edge is crossed.  As the decaying trajectory reaches the outermost part of the spiral, in the second stage (red), the edge no longer brackets the trajectory from both sides but is only located on the side closer to the ``turbulent'' state.  Consequently the lifetime divergence now is only observed for $\lambda >1$. 

In the model the lifetime variations on rescaling can be directly inferred from the state space portrait in frame (a). To check for a similar structure of the unwrapping edge in the full system, where we do not have a full graphical representation of state space, we need to choose a specific direction in which we locate the edge relative to the decaying trajectory. Since the edge is of codimension-one any fixed generic direction will not be tangent to the manifold and thus allow detection of the relative location of edge to the decaying trajectory. We here follow exactly the same protocol as for the model system and use linear rescaling along a line connecting the point in state space and the laminar state, $\mathbf{u}_{\lambda} \left(t_0\right) = \lambda \mathbf{u}\left(t_0\right)$, which again corresponds to the directions indicated by dotted lines in figure \ref{fig:edgeround}(a). The lifetime variations in panel (c) indeed show the same features expected from the model. For a point early in the decay (green) sharply increasing lifetimes indicate the edge being located on both sides of the trajectory. (Note that we stop the time integration after 1000 units and therefore not resolve the divergence.) Later (red), the edge is only found on one side, the turbulent side ($\lambda>1$) of the trajectory. 
Thus, the relative location of the edge exactly corresponds to the green and red points along the model trajectory. The transition from an edge located on both sides of the trajectory to it being on only one side is consistently observed within decaying trajectories and may define the exit of the trajectory from the turbulent saddle. 

To support this interpretation of an unwrapping of the edge we need to further demonstrate that the part of the edge that was first lying on the ``laminar side'' early in the decay and then is found on the ``turbulent side'' is actually dynamically connected and thus the ``same'' part of the edge. In the reduced model we track the edge trajectories using bisection for the two initial edge pieces indicated by crosses in frame (b). The dynamics of $\|\mathbf{x}\|$ against time are plotted for the decaying trajectory (blue) and two edge pieces (green and red) in frame (d). The edge trajectory found on the turbulent side during the later decay stage (red) only shows a small additional oscillation and then closely follows (up to a time shift) the edge trajectory found on the laminar side in the late decay stage (green). Both trajectories thus very quickly  follow the same path indicating the direct dynamical connection of the parts of the edge (visualized as solid red line in the schematic panel (a).  
Equivalent edge tracking is carried out in pCf in frame (e), again for the edges marked by crosses in frame (c). 
Precisely as for the model, the dynamics of the edge pieces for plane Couette flow only initially differ. Then they are  smoothly converging to the same dynamics. This demonstrates a dynamical connection between the pieces of edge on the ``laminar'' and ``turbulent'' sides during transition and is consistent with trajectories decaying around the edge as demonstrated by the coloured edge parts of part (a). This unwrapping happens for turbulent energies and thus very early in the decay. Consequently, a decaying trajectory does not pass close to the edge state whose guiding role for the decay remains questionable. 

\begin{figure}
\center\includegraphics[width=0.85\textwidth]{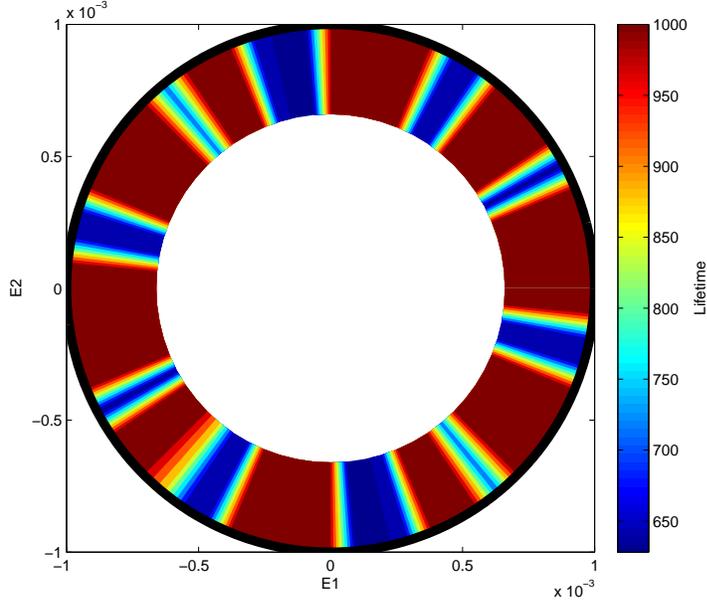}
\caption{Evidence of the edge connecting to an upper branch solution of pCf. Lifetime of initial conditions generated by perturbing a solution embedded in the turbulent saddle. Perturbations lie on the circle of radius $10^{-3}$ (black curve) in the (E1,E2)-plane where E1 and E2 are the two leading unstable eigenvectors of the solution. Time integration is limited to 1000 time units. We observe both trajectories indicating smooth decay to the laminar state (blue) and those who undergo extended turbulent dynamics (red). Between these regions lie critical angles whose initial conditions lie on the edge. The perturbation level used is within the linear regime for evolution away from the upper branch solution, demonstrating that the edge not only gets very close to the upper branch but the solution is actually embedded within it.
}
\label{fig:circle2}
\end{figure}

\section{Lifetimes near the upper branch}

As a final test for the edge being dynamically connected to the turbulence supporting saddle and wrapped around it, we focus on the state space region around upper branch solutions. In the model the upper branch state is interpreted as corresponding to turbulence and bifurcates in a fold-bifurcation with the edge state. The heteroclinic connection from upper branch to lower branch forms the edge. In pCf the turbulence supporting saddle contains many invariant structures \citep[see][]{kawahara2001,gibson08}. One of the embedded solutions associated with the turbulent saddle is the upper-branch Nagata equilibrium that is created together with the edge state. For this specific state within the turbulent saddle we study whether the edge wraps around it. Following \cite{gibson08} we capture the dominant dynamics close to the upper branch Nagata solution by projecting the dynamics onto the plane spanned by the two leading unstable eigenvectors E1 and E2. Considering initial conditions on a circle of radius $10^{-3}$ in the (E1,E2)-plane around the upper branch (where linearized dynamics govern the initial evolution) we compute the lifetimes to detect the edge. In figure \ref{fig:circle2} the lifetime is shown as a function of the angle coordinate. Abrupt jumps from short (blue) to long (red) lifetimes demonstrate that the edge can indeed be located at multiple points within a distance of $10^{-3}$ from the upper branch Nagata solution. 
Finding points of the edge within a distance governed by the linearized dynamics indicates that the edge not only gets very close to but indeed dynamically connects to the upper branch Nagata solution. Thus, the edge is indeed dynamically connected to the upper branch solution embedded in the turbulent saddle and thereby to the saddle itself. Additional support for this interpretation comes from recent studies by \cite{kreilos-periodic} in a smaller geometry. As the Reynolds number is increased, the turbulent saddle is shown to emerge due to a collision of a chaotic attractor with the stable
manifold of the state that after the collision turns into the edge state. Thus, the dynamical connection of the chaos supporting structures and the edge we observe at turbulent $\Rey$ is established already in the formation of the saddle.
 
Together these observations clearly demonstrate that the the edge of chaos is not separated from the turbulent saddle but wraps around the invariant solutions embedded in it. This provides further evidence that the low-dimensional model indeed captures the fundamental structure of pCf state space.

\section{Conclusions}

We have quantified the decay of turbulence in a periodically continued plane Couette cell with more than $10^5$ degrees of freedom. The decay proceeds in two steps which, for $\Rey=380$ presented here, each last about 250 time units. In the initial phase the downstream vortices disappear and the field loses its three dimensional structure, breaking the feedback cycle. In the second phase the remaining streak patterns are viscously damped away. Ensemble averages indicate that there are no distinct preferred decay paths, rather a continuum of routes and decay rates with considerable variation. Specifically, there is no evidence for a specific guiding role of the edge state. 
The absence of distinct decay paths can hardly be reconciled with the traditional view of a codimension-one edge of chaos that globally separates turbulence from laminar dynamic and that - in order to allow for the observed spontaneous turbulence decay - might be traversed at specific portals. Instead the observations are fully compatible with an edge that is not extended to infinity in all directions of state space but dynamically connected and wrapped around the turbulent saddle. Monitoring the location of the edge relative to a decaying trajectory we provide direct evidence that the ``unwrapping'' during decay found in a 2D model by Lebovitz is also the process in plane Couette flow. Finally we demonstrate the presence of the edge in an area of state space associated with turbulence, namely close to an upper branch solution.

In conclusion, an edge of chaos that is wrapped around and dynamically a part of the turbulent saddle appears to be the correct scenario not only for low-dimensional models but full shear flows at moderate Reynolds numbers, where spontaneous turbulence decay is observed despite the presence of a codimension-one edge of chaos. 

The authors would like to thank Norman Lebovitz, John F. Gibson and all of those involved in the 2011 WHOI GFD Summer program funded by OCE-0824636 and N00014-09-1-0844. MC was an GFD fellow and is funded by the EPSRC. TMS acknowledges support by the NSF, through grants DMS-0907985 and DMR-0820484, the Kavli Institute for Bionano Science and Technology at Harvard and the Max Planck Society.

\bibliographystyle{jfm}
\bibliography{paper}

\end{document}